\documentclass[12pt,final,twocolumn]{IEEEtran}
\usepackage{epsfig,endnotes}
\usepackage{amsfonts}
\usepackage{cite}
\usepackage{graphicx}
\usepackage{algorithm}
\usepackage{algorithmic}

\begin{document}

\title{\Large \bf Balanced Overlay Networks (BON): An Overlay Technology for Decentralized Load Balancing}

\author{ Jesse~S.A.~Bridgewater, 
Vwani~P.~Roychowdhury
and P.~Oscar~Boykin,~\IEEEmembership{Member,~IEEE}
\thanks{Jesse~S.A.~Bridgewater and Vwani~P.~Roychowdhury are with the
Electrical Engineering Department, University of California, Los Angeles, CA
90095}
\thanks{P.~Oscar~Boykin is with the Department of Electrical and Computer
Engineering, University of Florida, Gainesville, FL 32611}}
\date{}

\maketitle
\begin{abstract}
We present a novel framework, called balanced overlay networks (BON), that provides scalable, decentralized load balancing for distributed computing using large-scale pools of heterogeneous computers. Fundamentally, BON encodes the information about each node's available computational resources in the structure of the links connecting the nodes in the network. This distributed encoding is self-organized, with each node managing its in-degree and local connectivity via random-walk sampling. Assignment of incoming jobs to nodes with the most free resources is also accomplished by sampling the nodes via short random walks. Extensive simulations show that the resulting highly dynamic and self-organized graph structure can efficiently balance computational load throughout large-scale networks. These simulations cover a wide spectrum of cases, including significant heterogeneity in available computing resources and high burstiness in incoming load.  We provide analytical results that prove BON's scalability for truly large-scale networks: in particular we show that under certain ideal conditions, the network structure converges to Erd{\"o}s-R{\'e}nyi (ER) random graphs; our simulation results, however, show that the algorithm does much better, and the structures seem to approach the ideal case of d-regular random graphs. We also make a connection between highly-loaded BONs and the well-known ball-bin randomized load balancing framework.

\end{abstract}
\section{Introduction}
\begin{figure*}
\begin{center}
\includegraphics[scale=0.91]{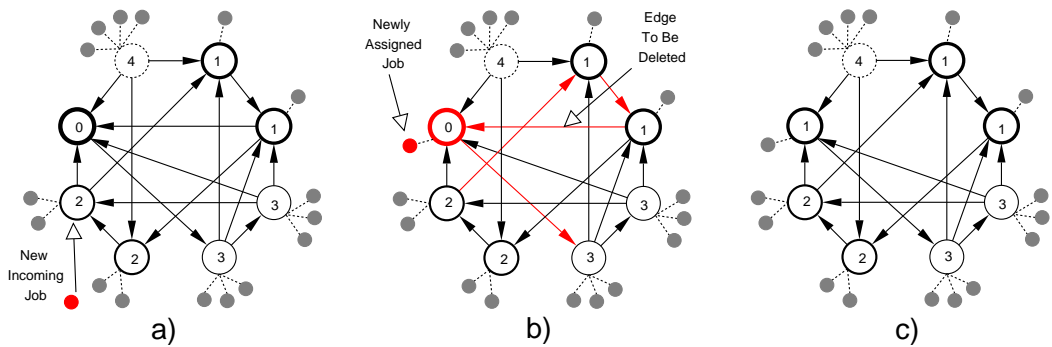}
\caption{New jobs are assigned by using a greedy random walk. The large nodes
depict computers in a schematic BON network while the small filled nodes are
jobs running on the node to which they are connected. The label for each of
the computer nodes denotes the current number of jobs it is running. Part {\bf a)} shows new load entering the network. In {\bf b)} we see that the node where load arrives initiates a random walk which keeps track of the degree(free resources) of each visited node. The largest degree node(most free resources) is selected to run the new load. To compensate for the additional load, the node which accepted the new load deletes one of its incoming edges to account for its diminished resources. The resulting network is depicted in {\bf c)}.}
\end{center}
\label{bon_diag_start}
\end{figure*}

\begin{figure*}
\begin{center}
\includegraphics[scale=0.91]{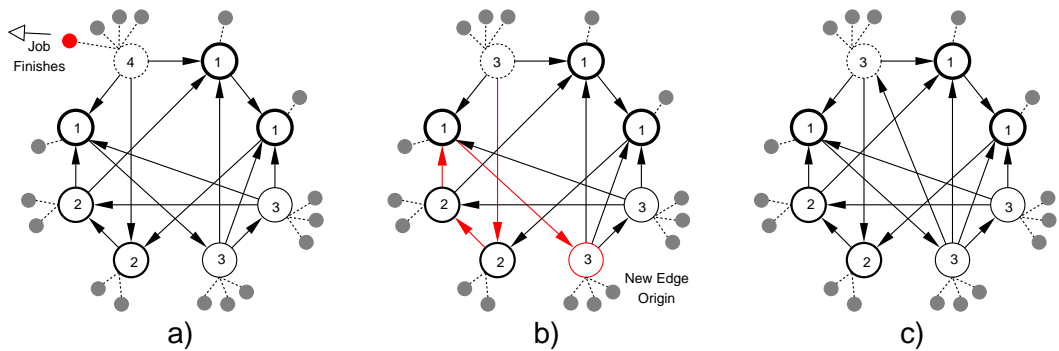}
\caption{When a running job finishes, the host node may need to increase its connectivity to advertise its increased resources. Subpart {\bf a)} shows a job finishing and thus leaving the network. In {\bf b)} the node where load finishes initiates a random walk. The last node on the walk will be the origin of a new edge incident on the walk initiator as seen in {\bf c)}. This new edge represents the increase in available resources on the node where the job just completed.}
\end{center}
\label{bon_diag_end}
\end{figure*}

Distributed computing was one of the earliest applications of computer
networking and many different methods have been developed to
harness the collective resources of networked computers. Some important
architectures include centralized client-server systems, DHT-based
systems, and diffusive algorithms.  
Here we introduce the concept
of balanced overlay networks (BON) which takes the novel approach of encoding
the resource balancing algorithm into the evolution of the
network's topology. Each node's in-degree is kept proportional to its unused
resources by adding and removing edges when resources are freed and consumed
as depicted in Figs. \ref{bon_diag_start} and \ref{bon_diag_end}. As we will show, 
this topology makes it possible to efficiently locate nodes with the most free
resources, which in turn enables load balancing with no central server.

This work makes several novel contributions to distributed
computing and resource sharing. First, BON is decentralized and
scalable with known lower bounds on balancing performance. While other
decentralized  load-balancing algorithms (e.g., Messor; see also
Section \ref{sec:related_work} for more detailed comparisons)  have been proposed in the
literature, performance and scalability analyses
for such algorithms, which guarantee almost-optimal performance as the
number of nodes becomes very large,
have been lacking.
Under certain ideal conditions, we show that
the network structure converges to a random graph that
is at least as regular and balanced as Erd{\"o}s-R{\'e}nyi (ER)
graphs.  Secondly, the algorithms and protocols for both network
maintenance
and job allocation are based only on local information and actions:
each node decides
the amount of resource or computing power it wants to share, and it
embeds this information
into the network structure via short random walks; similarly, jobs are
distributed based only on
information available through local explorations of the overlay
network. Thus, BON
is a truly self-organized dynamical system. Thirdly,
since the BON algorithm produces dynamic random
graph topologies, these resulting networks are very resilient
to random damage and also have no central point
of failure. Finally, we make a connection between the performance
of BON in some regimes with ball-bin random
load balancing problems \cite{mm:2001:twochoice}.  

It is also important to note that BON is
a novel paradigm for
for resource sharing of any kind and its applicability is not limited
to only distributed computing. The
in-degree of a node can be made to correspond to any type of shareable
resource. Then one can exploit the fact that
BON networks have low diameters
associated with random graphs, which makes
them easy to sample using short random walks.
Extensive simulation results support the efficacy of this approach in networks with a wide 
range of resource and load distributions. These simulations show that the
actual performance of the algorithm far exceeds the lower bounds mentioned
above.

BON is a very simple, realistic and easily implementable algorithm using standard 
networking protocols. The completely decentralized nature of the algorithm
makes it very well-suited to massive applications encompassing very large ensembles 
of nodes.  The following are a few examples of applications for which BON is
very well suited.

\noindent{\bf Single-System Image (SSI) LAN/WAN clusters}:
BON can be used for single-system image (SSI)
clusters in the same way that Mosix\cite{ab:1998:mosix} is used but without the need 
for all nodes to be aware of each other as is the case in Mosix. This can
allow BON to scale to very large system sizes.\\
\noindent{\bf Public Resource Computing}:
BON is also applicable to @HOME-style 
projects\cite{da:2004:boinc}.  This projects are typically special purpose for
each application. The decentralized nature of BON will allow multiple 
projects to share the same pool of computers.\\
\noindent{\bf Grid Computing}:
BON also has the potential to be integrated with
GRID\cite{if:2001:grid,sa:2005:invigo} systems for 
efficient resource discovery and load distribution across virtual
organizations (VOs).\\
\noindent{\bf Web Mirroring}:
Distributed web mirroring is an example of a non-computational application of
the BON algorithm. The system could allow a huge number of software users to
participate in providing download mirrors.

This paper is organized as follows.
Section \ref{sec:related_work} describes prior related load balancing research.
Section \ref{sec:bon_arch} introduces the BON architecture.
Section \ref{sec:theory} discusses theoretical analysis of BON's scalability.
Section \ref{sec:simulations} provides a description of the simulation setup and results.
Finally Section \ref{sec:practical} deals with practical considerations for implementing BON.

\section{Related Work}
\label{sec:related_work}
The authors have previously considered topologically-based load balancing
with a simpler model than BON which is amenable to analytical study\cite{jb:2005:statmech}. In that work each node's
resources were proportional to in-degree and load was distributed by performing
a short random walk and migrating load to the last node of the walk; this
method produces Erd{\"o}s-R{\'e}nyi (ER) random graphs and exhibits good load-balancing
performance. As we demonstrate in the current work, performing more complex
functions on the random walk can significantly improve performance.

The majority of distributed computing research has focused on
central server methods, DHT architectures, agent-based systems, randomized
algorithms and local diffusive techniques\cite{mt:1989:idle,do:2004:scalable,am:2002:messor,mm:2001:twochoice,rs:1994:diff,re:2003:token,bg:1999:load}. 
Some of the most successful systems to date \cite{da:2004:boinc,
ml:1988:condor-hunter} have used a centralized approach.  This can be
explained by the relatively small scale of the networked systems or by special
properties of the workload experienced by these systems. However since a
central server must have $O(N)$ bandwidth capacity and CPU power, systems that
depend on central architectures are
unscalable\cite{rl:1993:scalable,ok:1992:scalable}. Reliability is also a
concern since a central server is a single point of failure. BON
addresses both of these issues by using $O(log N)$ maximum communications
scaling and no single points of failure. Furthermore since the networks created
by the BON algorithm are random graphs, they will be highly robust to random failures.

The Messor project\cite{am:2002:messor} in particular has the same goal as BON
which is to provide self-organized, distributed load balancing. The
agent-based design of Messor also involves performing random walks on a
network to distribute load. However BON is designed specifically to reshape
the network structure so it can be efficiently sampled. Messor was inspired by
the notion of a swarm of ants that wander around the network picking up and
dropping off load.  Thus it is not clear how long the ant agents will need to
walk while performing the load balancing. It is the focus on topology that
distinguishes BON from other similar efforts. BON endeavors to reshape the
network topology to make resource discovery feasible with $O(\log N)$ length 
random walks. A simplified version of BON can be analyzed and thus 
we can put performance bounds on its behavior. Messor, while very intriguing, 
provides no analytical treatment.

Within the large body of research some techniques have been implemented including Mosix, Messor, BOINC, Condor, SWORD, Astrolabe, INS/Twine, Xenosearch
\cite{ab:1998:mosix, am:2002:messor, da:2004:boinc, ml:1988:condor-hunter, do:2004:scalable, vr:2001:astrolabe, mb:2002:instwine, ds:2003:xenosearch} and
others. Many of
these systems focus on providing a specific desired level of service for jobs.  
This contrasts to the approach taken by BON, Mosix and others in which processes are migrated to nodes where they will have the most
resources applied to them rather than a specific level of resources. The other systems are mostly based on DHT
architectures and provide for querying based on arbitrary node attributes and
link qualities. For complex distributed applications where each participating
node must have a certain level of resources and where the connectivity between
the nodes must have prescribed latencies, these DHT systems will
be the most suitable platform.  For many types of parallel scientific
computing however, BON's objective of placing a job where it will finish as quickly 
as possible is appropriate and desirable.

BON is designed to be deployed on extremely large ensembles of nodes. This is
a major similarity with BOINC\cite{da:2004:boinc}.  The Einstein@home project which processes gravitation
data and Predictor@home which studies protein-related disease are based on 
BOINC, the latest infrastructure for creating public-resource computing projects. Such projects are single-purpose and are designed to handle massive, embarrassingly parallel problems with
tens or hundreds of thousands of nodes. BON should scale to networks of this
scale and beyond while providing a dynamic, multi-user environment instead of
the special purpose environment provided by BOINC.

\section{The BON Architecture}
\label{sec:bon_arch}

\begin{figure*}
\begin{center}
\includegraphics[scale=0.91]{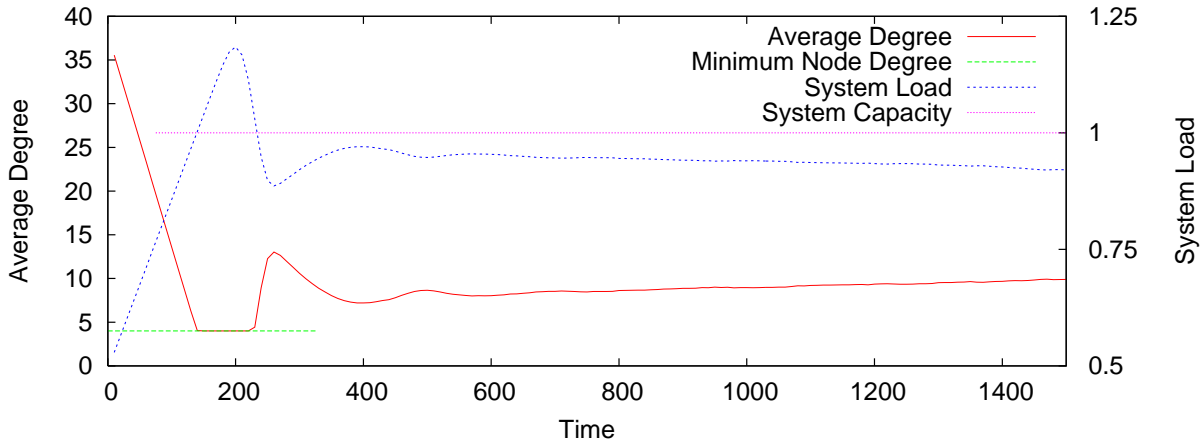}
\caption{The relationship between load and node degree is the basis for the
BON algorithm; a node with high in-degree is more likely to be visited on a
random walk and thus more likely to be the recipient of new load than a node
with low in-degree. As the total load 
increases, $\langle k\rangle$ decreases until the load becomes clipped. 
In the load clipped regime the algorithm remains the same but the mechanism behind 
the performance changes to becomes a ball-bin load balancing problem with 
$\log N$ choices. This change is due to the fact that there is no longer a connection
between free resources and in-degree.}
\end{center}
\label{load_degree}
\end{figure*}

\subsection{BON Topology}
The concept underlying BON is that the load characteristics of a distributed
computing system can be encoded in the topology of the graph that connects the
computational nodes.

In schematic terms, an edge in a BON graph represents some unit of unused
capacity on the node to which the edge points. Consequently when a node's
resources are being exhausted, its in-degree will decline as seen in Fig. \ref{bon_diag_start}.  Conversely when a
node's available resources are increasing, its in-degree will rise as seen in
figure \ref{bon_diag_end}.

Formally a BON is a dynamic, directed graph, $D = (E,V)$, where each node
$v_i \in V$ maintains $k^{(min)} \le k_i(t) \le k^{(max)}_i$ incoming edges. The maximum
incoming edges that a node can have, $ k^{(max)}_i$, is proportional to the
computational power of $v_i$.  
Each node, $v_i$, has a scalar metric $s_i(t)$ which is kept inversely proportional to $k_i(t)$. 
As $s_i(t)$ changes with time, $v_i$ severs or acquires new incoming links 
to maintain the relationship. In the context of distributed computing, 
$s_i(t)$ is a scalar representation of the current load experienced by node 
$v_i$. This means that each node will endeavor to keep its in-degree 
proportional to its free resources or inversely proportional to its load. 
Idle nodes will have a relatively large in-degree while overloaded nodes will have a small 
in-degree. The total unloaded resources of a node are proportional to it's
maximum in-degree, $k^{(max)}_i$.

\subsection{BON Algorithm}
Each node's load, $s_i(t)$, can change as new load arrives in the network or when existing
work is done.  When new load arrives at $v_i$, a short random walk is
initiated to locate a suitable execution site. Contained in this random walk is a BON resource discovery message(BRDM) which stores the merit function information for the most capable node visited so far on the walk. The fact that random 
walks will preferentially sample nodes with high-degree motivates the
mapping of node in-degree to free resources. The simplest approach to choosing a node
on the walk is to select the last node inserted into the BRDM. This case has been previously
explored\cite{jb:2005:statmech}. While this simple approach can be studied
analytically, simulation results indicate that large improvements to the 
balancing performance are possible by always keeping the most capable node's information.

Instead of performing a simple random walk and selecting the last node to
receive incoming load, the node on the walk with the largest power per load will be
the target (see Algorithm \ref{pick_target_alg}). Due to the mapping between load and in-degree this greedy random 
walk selects the least loaded node on the walk to receive new load which is the same as choosing the highest degree node when the network is not load clipped. This clipping occurs when a node has the minimum allowable in-degree. 
We will discuss the case when the network is load clipped in Section \ref{sec:ball-bin}. 
For time sharing systems the concept of overloading is not well-defined since a node with $L$ jobs will apply $1/L$ of its resources to each job. In the context of the BON algorithm load clipping simply means that nodes have the minimum
allowable in-degree and thus are no longer balancing load based on preferential
sampling. In practice a node in the clipped regime will be under very heavy computational load, but fundamentally it can still accept new jobs.

\begin{algorithm}
\caption{PickTarget(source): A new job entering the network initiates a random walk 
that maintain information about the node on the walk where the job would run the most
quickly.  The job is then assigned to that node when the walk is complete.}.
\label{pick_target_alg}
\begin{algorithmic}[1]
\STATE $ttl \leftarrow c \log N$, $v \leftarrow source$, $hops \leftarrow 0$
\STATE $obj_{max} \leftarrow \frac{v.Power()}{v.Load()+1}$, $target \leftarrow v$
\WHILE{$hops < ttl$}
  \STATE $v \leftarrow RandomOutNeighbor(v)$
  \STATE $obj_{temp} \leftarrow \frac{v.Power()}{v.Load() + 1}$
  \IF{$obj_{temp} > obj_{max}$}
    \STATE $obj_{max} \leftarrow obj_{temp}$
    \STATE $target \leftarrow v$
  \ENDIF
  \STATE $hops \leftarrow hops+1$
\ENDWHILE
\RETURN $target$
\end{algorithmic}
\end{algorithm}

\section{Analysis}
\label{sec:theory}
The performance of BON walk selection will be bounded below by the
performance of the standard walk selection.  Therefore although we do not
present a calculation of the load distribution for BON graphs, we can state
that it has the same scalability as the standard walk case described below.

The exact BON algorithm is difficult to analyze, however it is possible to 
place bounds on the balancing performance by simplifying the load distribution protocol.
We also calculate the bandwidth used by the algorithm and compare it to
a centralized model.

\subsection{Scalability}
The BON algorithm is difficult to study analytically due to the way in which
the random walks are sampled. However prior results\cite{jb:2005:statmech} 
show that a modified BON is more amenable to analysis.

Rather than selecting the node on the BRDM walk that can process an incoming job the fastest, 
one can simply select the last node of the walk.
In this model the average number of absent edges,$J$, in the $N$-node graph is identified as the total
number of jobs running. The maximum number of incoming edges that a node can
have will be called $C$ and the number of incoming edges to a given node is
denoted as $i$.
For the case when the average number of jobs remains
constant we can describe this system as a simple Markov process with
state-dependent arrival and service rates; it can be denoted by the standard
queueing notation as $M/M/\infty//M$. The arrival rate of new jobs is
proportional to the free resources,$i/(NC-J)$, of each node since jobs arrive
preferentially based on in-degree. Assuming that jobs terminate uniformly
randomly, the departure rate is $(C-i)/J$. Solving the birth and death Markov
process we obtain for the degree distribution:
\begin{eqnarray}
p_n = {C \choose n}\left[1-\frac{J}{NC}\right]^n \left[\frac{J}{NC}\right]^{C-n}
\end{eqnarray}

Defining the normalized load as $\alpha=J/NC$, the binomial distribution means that
for each node, each unit of capacity is occupied
with probability $\alpha$.  If $C=N-1$, this model recovers
the degree distribution for ER graphs:
\begin{equation}
p_n={N-1 \choose n}\left[\frac{E}{N(N-1)}\right]^n\left[1-\frac{E}{N(N-1)}\right]^{C-n}
\end{equation}
Where $E=N(N-1)-J$.

For a non-clipped network with uniform resource distribution, the variance of the degree 
distribution maps directly onto the variance of the load balancing.  This is
because each incoming edge represents free resources.  In a perfectly
balanced network, each node will have the same free resources. This ideally
balanced network would be a regular graph and thus the variance of the degree
distribution would be zero. For the simple case mentioned above the degree
distribution is binomial and thus it has a small but non-vanishing variance.

When the highest-degree (most free resources) node on a random walk is
selected to receive incoming load, that node's resources must be greater than
or equal to the resource of the last node on the walk.

In addition to this queueing model, it has been shown by information theoretic 
arguments\cite{jb:2005:statmech} that the simplified rewiring protocol described here
creates ER random graphs.

\subsection{Communications Complexity}
An important metric of performance for distributed computing is the network bandwidth required for a protocol.  It is clear that the architecture that requires the least total bandwidth is a central server. However the maximum bandwidth that any node must consume in a central system will not be the least. And while the total consumption of bandwidth is important, the bandwidth that any single node consumes can be a significant bottleneck for large central networks. Below we compare the bandwidth required by a centralized algorithm and by BON.
\subsubsection{Centralized}
The simplest non-trivial centralized architecture for a computing network is
the case where initially the central node, denoted $C$, knows the power and
load of each of the $N$ nodes that it controls. When a job on one of the nodes completes, that node will notify $C$ so that it can update its load state
information for the network. Obviously $C$ keeps track of assignments of new 
load to each of the nodes. This method does away with the need to periodically
probe every node in the network, however it is clear that the bandwidth, memory and CPU cost
that $C$ has to bear is still $O(N)$.  
Further assume a steady-state network load and that in every time unit, 
$N \beta$, jobs begin and the same number terminate. We further assume that
all the jobs will start at one of the computational nodes and that they will
then be sent to $C$ for assignment.
Now assume that for every job that is started a
relatively large $A$-byte packet, including the size of the program code and input data, 
must be sent from $C$ to $N_i, \quad i\in\{1,\cdots,N\}$ and that relatively
small $L$-byte
packets must be sent to the central server in response to changes in load. Therefore $C$ must
send $N \beta A$ bytes per unit time which consumes kernel resources and requires
 bandwidth that increases with $N$. The total bandwidth consumed by the entire
 centralized network is
\begin{equation}
B^{(C)}_T = N \beta \left[ A + L \right].
\end{equation}

This is also the same amount that $C$ must consume since it is involved in
every communications round. For $N_i,\quad \forall i \in \{1,\cdots,N\}$ the
bandwidth consumed will be $B^{(i)}_T = \beta \left[ A + L \right]$ which is
$O(1)$.

\subsubsection{BON}
For the decentralized BON algorithm, the network topology is now more complex
than for the central server.  While the graph of the central model was a star,
BON will look approximately like a random regular graph. Initially we will
assume that we begin with a correctly-formed BON.  As with the central model
we assume that $N \beta$ jobs begin and end at random nodes in each time unit.
Since there is not a central server, each node that initiates a new job
must send a BRDM to find a node to run the job. Every node
on the walk will need to replace the value of $obj =\frac{Power()}{Load()+1}$, ($L$ bytes of data), 
in the BRDM if $obj$ is larger than the objective function that is currently in the BRDM (see Algorithm \ref{pick_target_alg}. Since
the random walk will be $O(\log N)$ steps long, the total bandwidth of the
walk will be $B_w=L \log N$. Likewise when a job finishes, another
walk will be used to find a replacement for the removed edge. Factoring in the
cost of transmitting the program to the target node and needed handshaking
protocols, the total bandwidth consumed by BON is

\begin{equation}
B^{(BON)}_T = N \beta \left[ A + L\left\{ \log N +2\right\} \right]. 
\end{equation}

Therefore we can see that the total bandwidth cost of BON is $O(\log N)$
greater than the central model.  However a more important metric in many
situations will be the maximum bandwidth consumed by any of the nodes. In BON
each node will on consume bandwidth in proportion to how many jobs it
initiates and how powerful it is. Thus if all of the nodes use the network equally then each node
will consume $B^{(BON)}_T/N$ bandwidth, which is logarithmic in the size of the
network. This contrasts to the $O(N)$ bandwidth needs of the central server.

\section{Simulations}
\label{sec:simulations}
\subsection{Simulation Description}

For the simulations, each node in a BON network is a computer with power equal
to its maximum degree minus its minimum degree, $P_i=k_{i}^{(max)}-k^{(min)}$. One
unit of power can process a unit of load in each unit of time. Jobs run on these
computers in a time-sharing fashion with each of the $L$ jobs of a computer
equally sharing the node's power at each time step. The simulations
deal only with CPU power as the objective function of the balancing. Other
features such as memory ushering will not be simulated but will be added as
features in the reference implementation. Simulations of the BON system were 
performed using the Netmodeler package. Two type of experiments were done.  

The first experiments are very idealized using uniform node power, uniform job
arrival rates and Poisson-distributed job sizes. Equation \ref{easy_params}
indicates that all nodes have $k^{(max)}=71$, the size of each job is Poisson
distributed and that at each time step $\beta$ jobs are created. For different
simulations $\beta$ and $\nu$ will have different values in order to show a
wide range of system behavior. While this setup is very idealized, it might apply 
to cluster computing.

\begin{eqnarray}
P(k) & = & \cases{1, \quad k=71\cr
0, \quad k\ne 71\cr},\nonumber\\
P_{\nu}(j) & = & \frac{\nu^j e^{-\nu}}{!j},\nonumber\\
P_\beta(b) & = & \cases{1, \quad b=\beta\cr
0, \quad b\ne \beta\cr}
\label{easy_params}
\end{eqnarray}

The second type of simulation (Eqn. \ref{hard_params}) uses power-law distributions for all parameters
including job arrival rate, node power and job size. This configuration
represents a situation where every important system parameter is
distributed in a bursty, heavy-tailed way. Heavy-tailed distributions are common in many real systems\cite{mm:2003:powerlaw} including networks and thus these simulations provide a fairly realistic idea of how the system will perform under real loads. Most importantly for simulation
performance the computing power ranges from $1$ unit of power to $300$ units
of power. This is at least ten times the range of performance seen in commonly
used CPUs. As we will revisit in the performance evaluation, having many nodes
that can only accept a few processes prior to being load-clipped will impact the
balance distribution simply due to quantization effects. This issue will have
design implications for the implementation.
\begin{eqnarray}
P_{k} & \propto & k^{-1}, \quad k^{(min)}+1 \le k \le 304 ,\nonumber\\
P_{j} & \propto & j^{-1}, \quad 32 \le j \le 1024 ,\nonumber\\
P_{b} & \propto & b^{-1}, \quad 1 \le b \le b_{max}
\label{hard_params}
\end{eqnarray}

In all of these simulations we begin with a randomly-connected network subject to the initial degree distribution.  However if one starts with a completely ordered network with $O(N)$ diameter, the graph will quickly converge to the low-diameter structure depicted in these simulations.

\subsection{Graph Structure} 
\begin{figure*}[t]
\begin{center}
\includegraphics[scale=0.91]{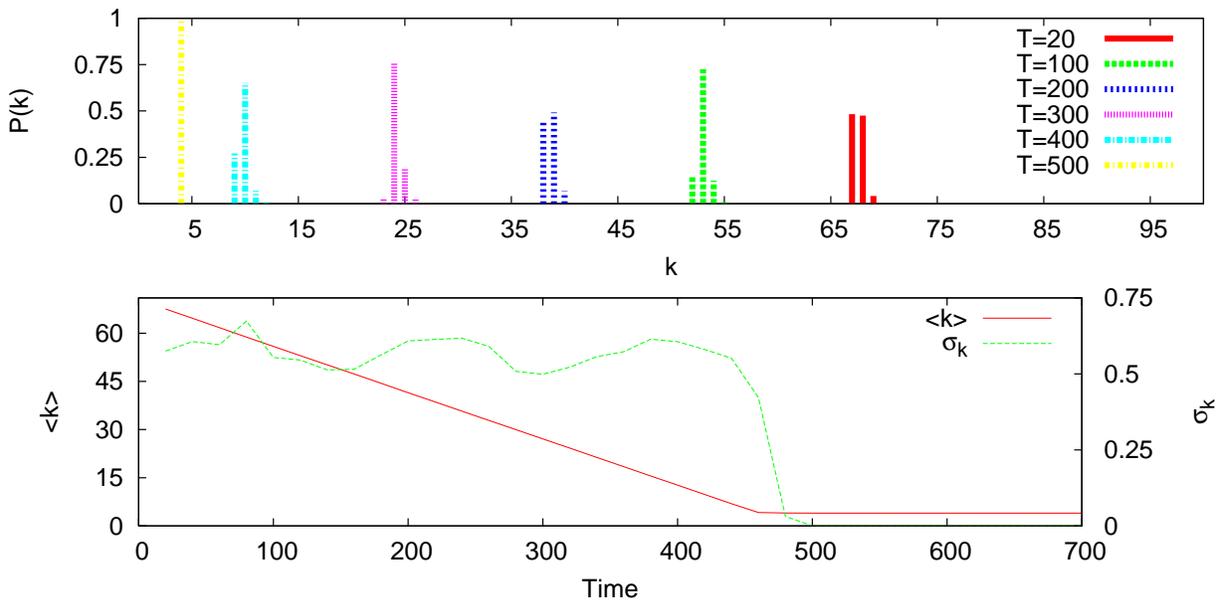}
\caption{When all nodes have the same resources,  BONs will be
approximately regular graphs. These degree distribution snapshots (top graph) 
of an evolving network that is getting progressively more loaded show the 
regular nature of the graph over a wide range of load conditions.}
\label{easy_graph}
\end{center}
\end{figure*}

\begin{figure}
\begin{center}
\includegraphics[scale=0.91]{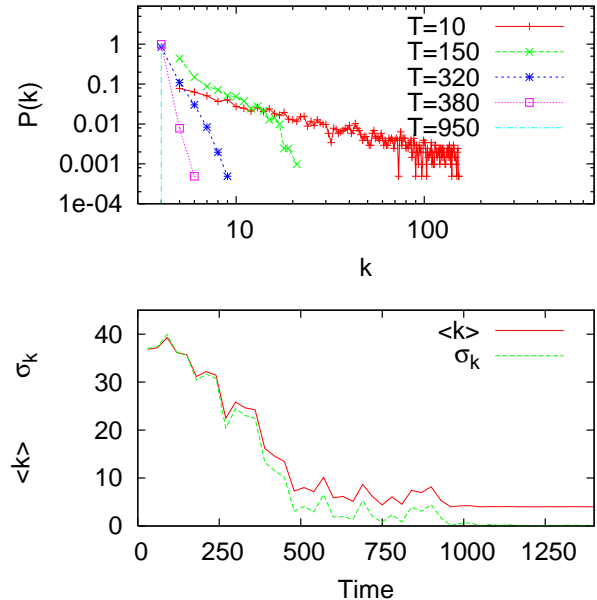}
\caption{For a network with a power-law resource distribution, the degree
distribution also has a power-law form with the exponent getting more negative
with increasing load.}
\label{hard_graph}
\end{center}
\end{figure}

The idea at the heart of BON is that the graph structure can capture the load
state of a computational network. 
In section \ref{sec:theory} we discussed prior theory results that describe the
structure of graphs formed using algorithms similar to BON. We now present
simulation results for both the uniform and heavy-tailed systems
described above. The degree distribution for a balanced overlay network
matches the resources of the constituent nodes for both uniform and power-law
resource distributions as seen in Figs.  \ref{easy_graph} and \ref{hard_graph}.
\begin{figure}[t!]
\begin{center}
\includegraphics[scale=0.91]{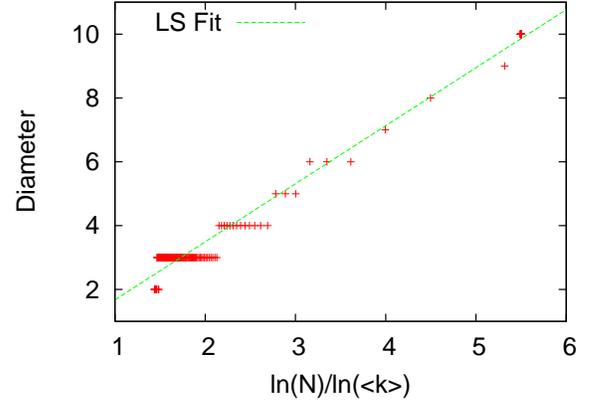}
\caption{
BON graphs obey the random graph scaling relationship between diameter and
average degree: $Diameter \propto \ln N / \ln\langle k\rangle$. This BON graph was generated using the uniform simulation parameters from Eqn. \ref{easy_params}.}
\label{easy_graph_diameter}
\end{center}
\end{figure}

\begin{figure}[b!]
\begin{center}
\includegraphics[scale=0.91]{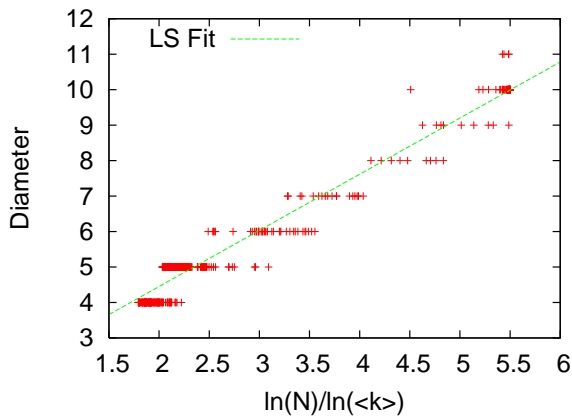}
\caption{
One can see that this system obeys the random graph scaling relationship
between diameter and average degree: $Diameter \propto \ln N / \ln\langle
k\rangle$. This BON graph was generated using the power-law simulation
parameters from Eqn. \ref{hard_params}}.
\label{hard_graph_diameter}
\end{center}
\end{figure}

\begin{figure}[t!]
\begin{center}
\includegraphics[scale=0.91]{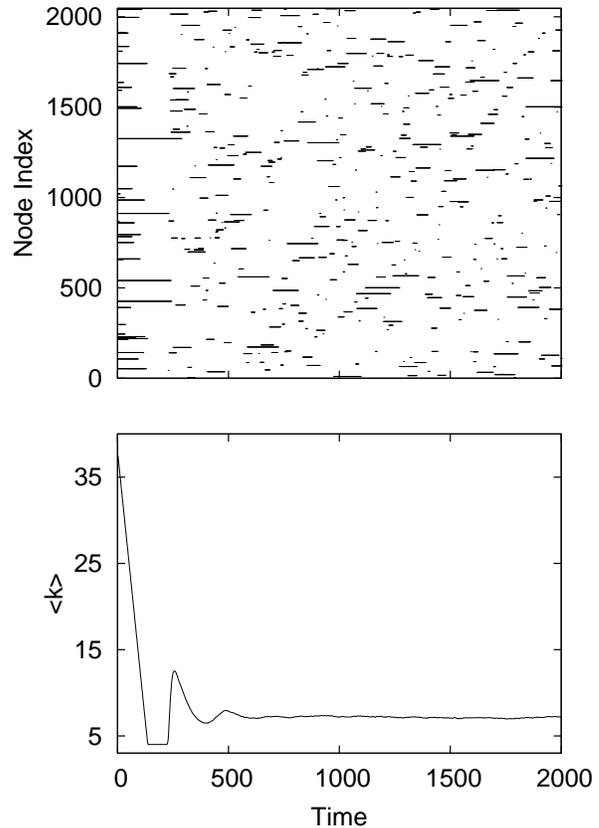}
\caption{As a BON network evolves there is significant turnover in
connections. In the top sub-figure the in-degrees of an arbitrary node, $v_0$, are depicted as a function of time. A vertical line, $t=a$, intersects zero or more points, $A_0$, which
is the set of nodes that have directed edges incident to $v_0$. This
illustrates that the structure of the graph changes significantly even when
the macroscopic properties such as average degree are not changing.}
\label{easy_graph_turnover}
\end{center}
\end{figure}

Figs. \ref{easy_graph_diameter} and \ref{hard_graph_diameter} show that BONs maintain
a low diameter and exhibit the property of random graphs that the diameter is
proportional to $\ln N/\ln\langle k\rangle$. The changes in connectivity can be seen in Fig. \ref{easy_graph_turnover}.

\begin{figure}[t!]
\begin{center}
\includegraphics[scale=0.91]{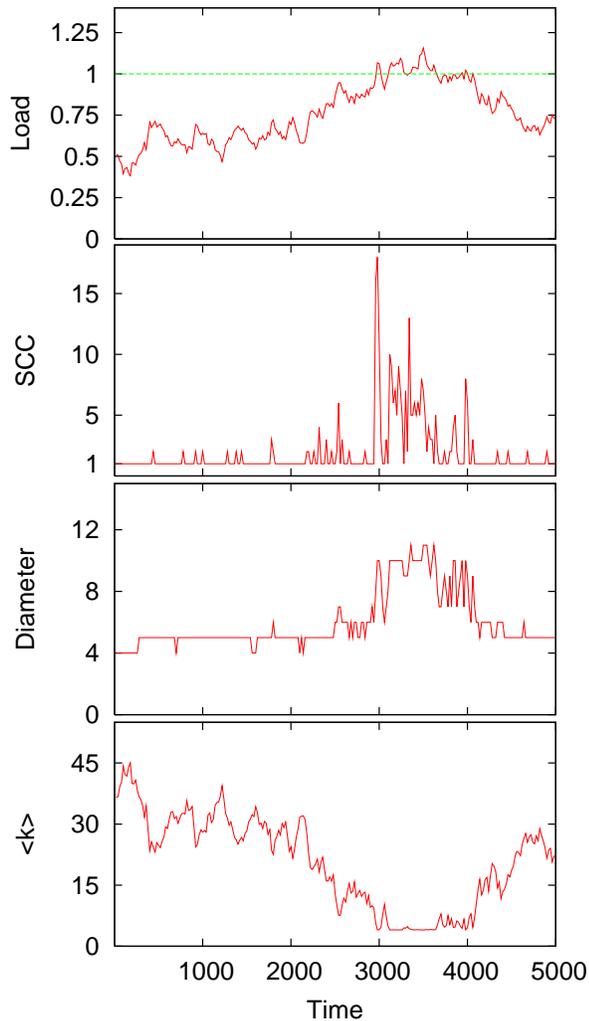}
\caption{In simulation we observe that BONs are always at least weakly connected
directed random graphs. This
$2048$-node BON with heavy-tailed parameters remains weakly connected even in the clipped regime. The number of strongly-connected components(SCC) does increase under
heavy load, but the number of SCCs returns to unity when the clipping condition passes.}
\label{hard_diameter}
\end{center}
\end{figure}

It is important that BON graphs remain at least weakly-connected as they
evolve. All simulations indicate that BONs remain weakly-connected, but that
when they are load-clipped they can acquire a complex strongly-connected
structure.  As the load surpasses $1$ (the clipping threshold) and the network becomes a $k^{(min)}$-regular graph, the number of strongly-connected components (SCC) increases. As shown in Fig. \ref{hard_diameter}, the number of SCCs falls back to unity when an overloaded network becomes less loaded. It is also important to note that the SCCs in an overloaded BON can change due to rewiring of the network. So while every node will not be able to communicate with every other node at each instant of time, the out-component of each node in the graph can change with time. Also the network does remain weakly-connected even when the network has many SCCs. 

\subsection{Load Balancing Performance}
When discussing load balancing performance we want metrics which
measure how closely load follows capacity.  When all the nodes are
equally capable, standard deviation is a convenient measure of balancing, when nodes are
heterogeneous, correlation coefficient is what we use.
\subsubsection{Simple Idealized System}

\begin{figure}[t!]
\begin{center}
\includegraphics[scale=0.91]{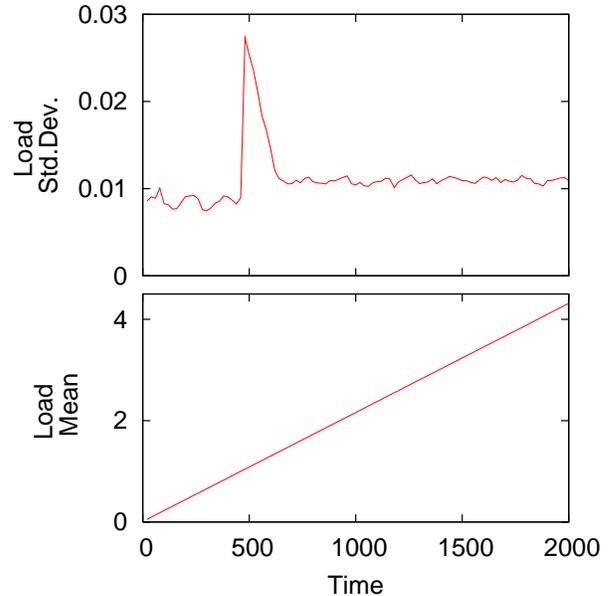}
\caption{This uniform resource, $2048$-node BON under increasing overload also shows that
the standard deviation of the load is low at about $1\%$. The difference
in performance as the network enters the clipped regime can also be
seen. At the clipping transition point the standard deviation experiences a spike which is likely due to jobs accumulating in a small SCC before additional rewiring can rebalance the load. After a short time this load imbalance dissipates as the rewiring allows load to be distributed throughout the network.}
\label{fig:easy_over_load}
\end{center}
\end{figure}

For the uniform simulation model, Fig. \ref{fig:easy_over_load} shows that the
ensemble standard deviation of the node load is just below $1\%$
when the network is in the under-loaded regime. When the network is clipped 
the standard deviation of the load is slightly higher than in the under-loaded
regime but still quite close to $1\%$. This difference in performance is
likely due to the transition from the degree-correlated load-balancing that is
in effect when the network is under-loaded to the ball-bin
load balancing that takes over when the network is clipped.

Another important measure of performance is how well BON performs in
comparison to a central system that places new jobs at the least loaded node
in the network. In the uniform configuration after $1000$ iterations the
central system has completed $501314$ jobs compared with $501238$ jobs being
completed by BON. This indicates that BON's job throughput is only about $0.01\%$ 
worse than the optimal schedule.

\subsubsection{Power-Law System}
\begin{figure}[t!]
\begin{center}
\includegraphics[scale=0.91]{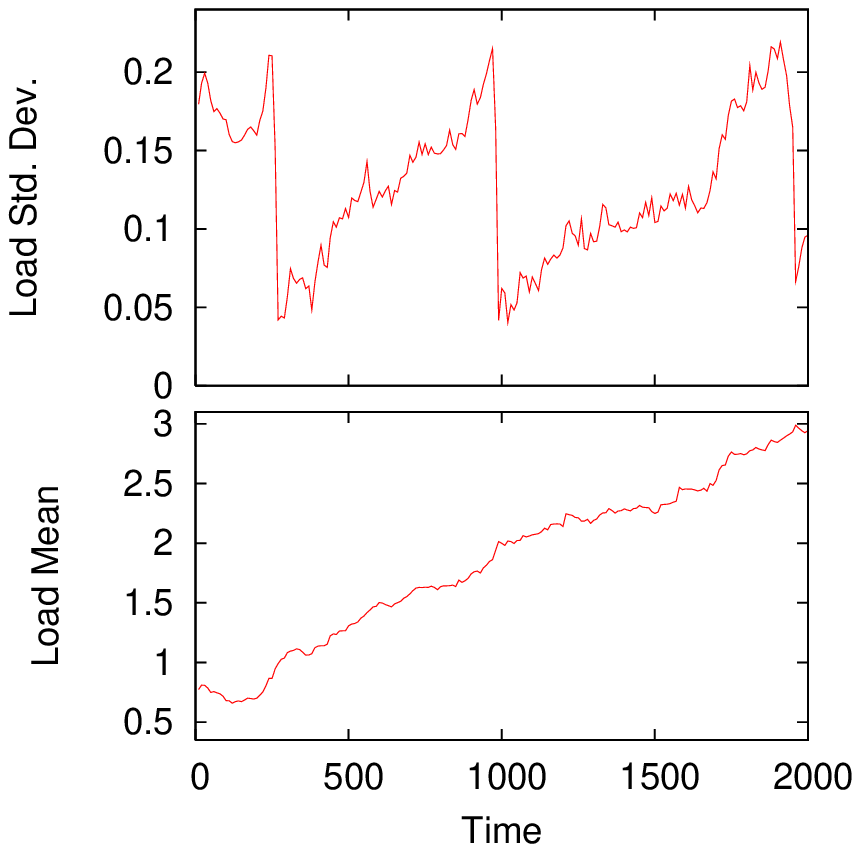}
\caption{Here we see the cyclical nature of the load standard deviation as a function of network load. The
dips happen when the network is at an integer multiple of the load clipping threshold. This
is due to the assumption in this model that most of the nodes have a very low
degree(power-law distribution). For instance if the network is $x\%$ loaded a
node with a maximum in-degree of $k_{min}+k_{load}$ will only be able to have
the same load as the rest of the network if $\frac{ J}{k_{load}} =
l_{load},\quad  \textrm{for all} J \in \{0,\cdots , k_{load} o\}, o\in
\mathbb{Z}^+ $.}
\label{hard_overload}
\end{center}
\end{figure}

\begin{figure*}
\begin{center}
\includegraphics[scale=0.91]{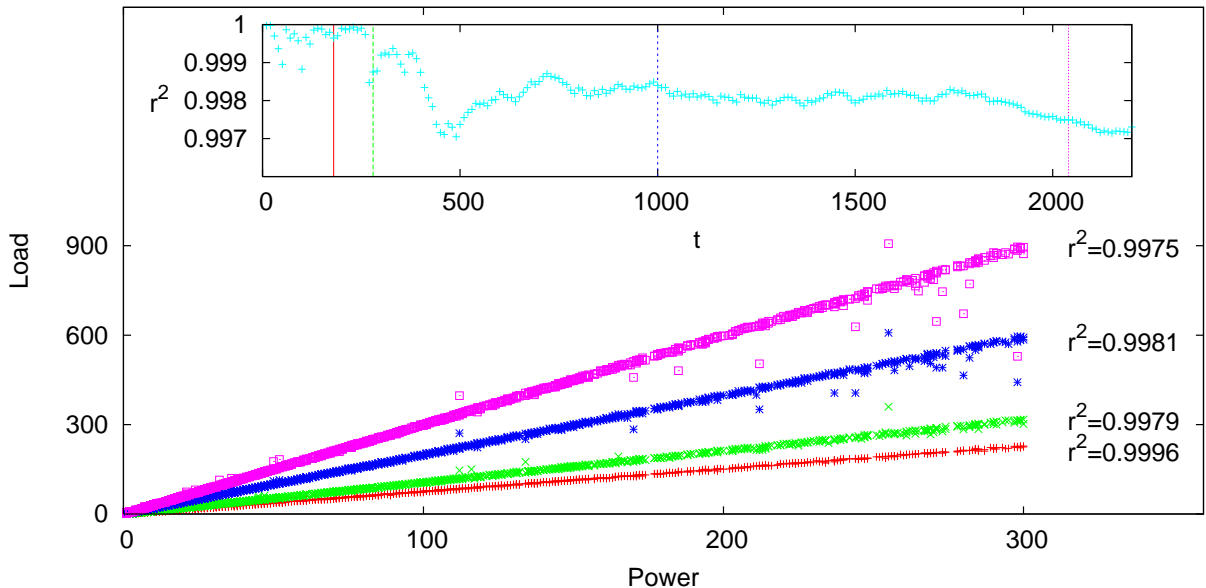}
\caption{For the case where we do have many nodes with a low maximum degree 
the correlation between node power in-degree is a more appropriate measure of 
performance than standard deviation.
and load. For a network that is getting increasingly loaded the load vs.
power is plotted at instants and the correlation is also calculated. Even when
the network is 3X the load clipping threshold, $r^2 > 0.99$.}
\label{hard_corr}
\end{center}
\end{figure*}

The power-law simulations illustrate an important design criterion for
practical implementations. For these simulations the power distribution of the
nodes is a power-law given in Eqn. \ref{hard_params}. The minimum power is $1$ and the
maximum power is $300$. Therefore there are many nodes that have very low power
resources. This means that for many values of the load it will be impossible
to get close to optimal balancing. For this reason the correlation between degree and free resources is used to evaluate performance as shown in Fig. \ref{hard_corr}. A good example is a node with $P=2$.
Because the load is defined to be $P/L$, where $L$ is the number of running
jobs, the load is limited to be non-negative integer multiples of $1/2$. Thus
if the network is $75\%$ loaded then this low-powered node is equally
unbalanced whether one or two jobs are running. By selecting a suitable
minimum power, one can bound this finite size effect.  For example if the
least powerful node has $P=5$ then it can always get within $10\%$ of the
optimal value. 
This finite size effect appears as cyclical behavior of the load standard 
deviation and can clearly be seen in Fig. \ref{hard_overload}.

As was done for the uniform simulation configuration, we compare centrally scheduled job throughput to BON throughput with
the same load trace. In the heavy-tailed configuration, after $1000$ iterations the
central system has completed $585872$ jobs compared with $585788$ jobs being
completed by BON. As with the uniform configuration, BON's job throughput in
the heavy-tailed configuration is only about $0.01\%$  worse than the optimal 
schedule. Please note that this result ignores the effects of job distribution 
latency on total throughput but it indicates that job placement is very close
to optimal when communications delays are ignored.

\subsection{Ball-Bin Regime}
\label{sec:ball-bin}

Every node in the graph must maintain a minimum degree to ensure that the
graph stays at least weakly-connected. For these experiments each node maintains at least 4
incoming edges which means that if the network's load becomes clipped then 
there is no longer a correlation between a node's degree and its
resources. For this reason the real metric that is sampled on the walk is the
amount of computing power that the next incoming process can expect on a 
given node. When the network is not clipped this is the same as
choosing the highest-degree node on the walk.  However for a clipped
network it selects the node on the walk that has the largest value of the expected power for the next incoming job as shown in algorithm \ref{pick_target_alg}.  Now consider that a
clipped network is approximately a regular random graph and thus a short
random walk will sample uniformly from the nodes in the network.  This problem
now shows itself to be very similar to ball-bin load 
balancing\cite{ya:2000:balanced,mm:2001:twochoice}.

\begin{equation}
\cases{\langle k\rangle  > k^{(min)}, & preferential sampling\cr
\langle k\rangle  = k^{(min)}, & ball-bin sampling\cr}
\label{algorithm_regime_eq}
\end{equation}

In ball-bin systems a ball is uniformly randomly assigned to one of $N$ bins.
As this process is repeated a distribution of bin population emerges and has
been studied extensively under many kinds of assumptions. The important result
from ball-bin systems is that if one probes the population of more than one
bin prior to assigning a ball, the population of the most full bin will be
reduced exponentially in $N$. This work is often referred 
to as the ``power of 2 choices''\cite{mm:2001:twochoice}. 

In the load-clipped regime we have a similar situation where instead of two
choices we have the power of $\log N$ choices. Each random walk on the
$k^{(min)}$-regular graph will sample uniformly randomly from the nodes. Then
the least loaded nodes from the $\log N$ choices will be the target to accept
the new load. This connection is made to give intuition for why BON
should function in the overloaded regime but we will not examine this aspect
of the system in detail here. Detailed followup work will be
performed to compare overloaded BON performance to the theoretical predictions
of ball-bin systems. 


\section{Practical Considerations}
\label{sec:practical}
\noindent {\bf 1) Network: }
In the presented simulations some networks have nodes with hundreds of
incoming edges. While modern computers can easily maintain hundreds of
simultaneous TCP connections, for BON it may be preferable to use UDP for
some aspects of the network.
BON nodes will interact with edges when load is distributed using random walks
and when edges are being created or destroyed. These edges are important because they maintain the state of the network, but if a connection goes down it can easily be replaced without affecting system performance. However when load distribution messages
are random walking through the network, reliable transmission is important. There are numerous ways to provide for reliable communications between nodes using both TCP and UDP. Efforts to use fast light-weight protocols while minimizing latency will be important design issues for a BON implementation.
Most connections at any given time will not be transmitting BRDMs but will be
maintaining the network state. For state maintenance the use of UDP will drastically reduce overhead
compared to TCP and will allow a much larger number of edges to be maintained with less overhead that
TCP. Using soft state information from packet traffic to perform keep-alive
operations will help mitigate connection maintenance overhead.

\noindent {\bf 2) State Encoding: }
For the load objective function we will follow a similar approach to the Mosix SSI cluster computing
system\cite{ab:1998:mosix}. 
The Mosix migration algorithm is heuristic in nature and
basically attempts to run processes where they will finish the most quickly.
Various historical data about the process execution and node load and
resources are used to judge which node can process a job with the least cost.  
Additionally Mosix uses a memory ushering protocol to migrate processes away 
from nodes with depleted memory resources. This ushering is done in favor of
trying to integrate the memory and CPU metrics into a single scalar value. These methods have been motivated by real system profiling and have proved to be successful.
Thus the node resource that will be kept proportional to in-degree is the
available CPU resources of the node. In particular we wish for new load to be
assigned to the node $v_i$, where $i=argmax_{j}\left\{
\frac{P_j}{L_j+1}\right\}$.  Here $P_j$ is $v_j$'s power which can be any
standardized way of representing the number of operations per unit time that a
node can perform and $L_j$ is the number of processes competing for
$P_j$ (UNIX load).  The details of how to weight integer, floating-point and
other processing characteristics will not be considered here but it will be
assumed that a reasonable benchmark of CPU performance can be constructed and 
run periodically on each BON node.

\noindent {\bf 3) Load Quantization: }
Since computing power is represented by the edges in the network, it is
important to scale the power that each edge represents in order to get the
most load balancing performance for the least bandwidth and state
maintenance.
The initial implementation of BON will specify a computer to be the baseline of
computer power. As computer performance changes, adaptive base-lining can be 
performed to automatically scale how much computing
power is represented by a BON edge. All other node powers are computed w.r.t.
the $k$th percentile of benchmarks. That is all nodes in the $k$th
percentile will have the baseline power of $P_b$ and will maintain at most
$k^{(b)}-k_{min}=5$ baseline resource edges.  All other nodes will maintain
\begin{equation}
 k_i-k_{min} = \frac{P_i(t)}{P_b} k^{(b)}
\end{equation} 
resource edges.
Choosing $k_i^{(max)}-k_{min} \ge 5,\quad \forall i$ ensures that even the least powerful nodes in the network 
can have a load that is within $10\%$ of optimally balanced.

\section{Conclusion}
\label{sec:conclusion}

Balanced overlay networks (BON) is a novel decentralized load-balancing
approach that encodes the balancing
algorithm in the evolving structure of the graph that connects the
resource-bearing nodes. BON is scalable, self-organized and relies only local
information to make job assitgnment decisions. New jobs are assigned to a node by a
random walk on the graph which not only samples the graph
preferentially, but also selects the highest-degree node that was visited on
the walk. Each node's unused resources are proportional to its degree so this 
approach works very well when a network is not loaded beyond its clipping point.  
When a BON is clipped the relationship between load and in-degree
breaks down but the balancing performance
remains quite good due to the so-called ``power of two choices'' in ball-bin
load balancing. Based on previous theoretical results and extensive simulation
results, BON is seen to be efficient and practical. Further ongoing work on this
problem includes geographical awareness extensions using more complex walk
objective functions, a reference implementation
on PlanetLab, theoretical analysis of the random walk with greedy node
selection, algorithmic optimizations and a full comparison of overloaded regime results with the
predictions of ball-bin random load-balancing. Finally it should be noted that this
is only one possible way to encode information about a network in its topology;
other distributed algorithms may benefit from using graph state to bias node
selection.

\bibliographystyle{IEEEtran.bst}
\bibliography{bon_conf_01}

\end{document}